\documentclass[aps,pra,twocolumn,showpacs,superscriptaddress,letterpaper]{revtex4}

\usepackage{graphicx}
\usepackage{amsmath}
\usepackage{amssymb}
\usepackage{dcolumn}
\usepackage{bm}

\begin{document}

%
%
%
%
%
\def\bra#1{\mathinner{\langle{#1}|}}
\def\ket#1{\mathinner{|{#1}\rangle}}
\def\braket#1{\mathinner{\langle{#1}\rangle}}
\def\Bra#1{\left<#1\right|}
\def\Ket#1{\left|#1\right>}
{\catcode`\|=\active
  \gdef\Braket#1{\left<\mathcode`\|"8000\let|\BraVert {#1}\right>}}
\def\BraVert{\egroup\,\mid@vertical\,\bgroup}
%

{\catcode`\|=\active
  \gdef\set#1{\mathinner{\lbrace\,{\mathcode`\|"8000\let|\midvert #1}\,\rbrace}}
  \gdef\Set#1{\left\{\:{\mathcode`\|"8000\let|\SetVert #1}\:\right\}}}
\def\midvert{\egroup\mid\bgroup}
\def\SetVert{\egroup\;\mid@vertical\;\bgroup}

\newcommand{\derfrac}[2]{\genfrac{}{}{}{}{d#1}{d#2}}

\title{Coherent Quantum Engineering of Free-Space Laser Cooling}

\author{Josh~W.~Dunn}

\affiliation{JILA, University of Colorado and National Institute of Standards and Technology, and
Department of Physics, University of Colorado, Boulder, Colorado 80309-0440}

\author{J.~W.~Thomsen}

\affiliation{The Niels Bohr Institute, Universitetsparken 5, 2100, Copenhagen, Denmark}

\author{Chris~H.~Greene}

\affiliation{JILA, University of Colorado and National Institute of Standards and Technology, and
Department of Physics, University of Colorado, Boulder, Colorado 80309-0440}

\author{Flavio~C.~Cruz}

\affiliation{JILA, University of Colorado and National Institute of Standards and Technology, and
Department of Physics, University of Colorado, Boulder, Colorado 80309-0440}

\affiliation{Instituto de Fisica Gleb Wataghin, Universidade Estadual de Campinas, CP. 6165,
Campinas, SP, 13083-970, Brazil}

\date{\today}


\begin{abstract}
We perform a quantitative analysis of the cooling dynamics of three-level atomic systems
interacting with two distinct lasers. Employing sparse-matrix techniques, we find numerical
solutions to the fully quantized master equation in steady state. Our method allows straightforward
determination of laser-cooling temperatures without the ambiguity often accompanied by
semiclassical calculations, and more quickly than non-sparse techniques. Our calculations allow us
to develop an understanding of the regimes of cooling, as well as a qualitative picture of the
mechanism, related to the phenomenon of electromagnetically induced transparency.
Effects of the induced asymmetric Fano-type lineshapes affect the detunings required for optimum
cooling, as well as the predicted minimum temperatures which can be lower than the Doppler limit for
either transition.
\end{abstract}

\maketitle

Laser cooling has led to tremendous achievements, enabling extremely precise
measurements (see, for example, Ref.~\cite{frequency_standards_proceedings_01}), the development
of highly accurate atomic clocks~\cite{takamoto2005}, and the production and manipulation of
quantum-degenerate gases~\cite{anderson1995obe}. The atomic internal structure greatly restricts
the species for which laser cooling can be applied successfully, and determines whether ultracold
temperatures (on the order of a few $\mu$K) can be achieved. The basic Doppler cooling
mechanism~\cite{lett1989om,metcalf_book_99}, based on momentum transfer from a near-resonant light
beam, has a minimum temperature (typically a few hundred $\mu$K) proportional to the scattering
rate (linewidth) of the atomic transition, and given by the balance between friction (cooling) and
diffusion (heating) in the atom-laser interaction. Lower temperatures can be achieved either by exploiting the multilevel hyperfine structure
(Sisyphus~\cite{cohen-tannoudji_89,chu_89}, Raman~\cite{chu_92}, VSCPT~\cite{aspect1989lcb}), by
modifying the atomic scattering (spontaneous emission) for example in a near-resonant
cavity~\cite{chan2003oce}, or, more recently, by using weakly allowed, narrow two-level optical
transitions in a second-stage for atoms that have been pre-cooled by other
means~\cite{katori1999mot,kuwamoto1999mot}. Here, we explore cooling by dissipative radiation-pressure forces~\cite{metcalf_book_99} in three-level systems under two-color
excitation. Similar schemes have been studied previously for trapped
ions~\cite{morigi_00,roos2000edg}. In the electromagnetically-induced-transparency
(EIT) regime, e.g.,
with a weak "probe", and a strong "dressing" laser, an effective two-level system is engineered
that allows sideband cooling of ions to the ground state of the trapping
potential~\cite{roos2000edg,zoller_94}.

\begin{figure}
\begin{center}
\includegraphics[width=3in]{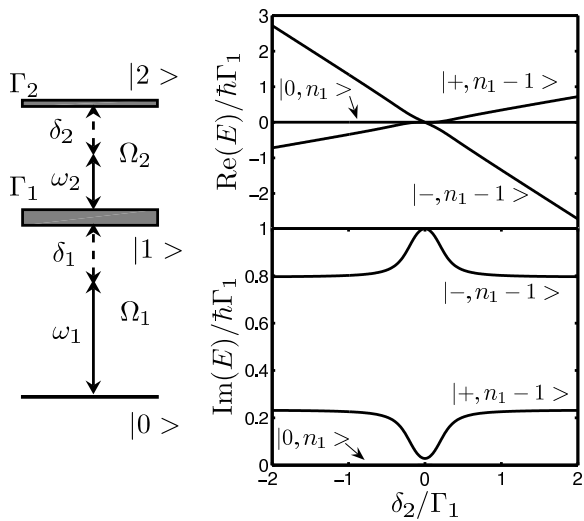}
\caption{\label{fig:cascade}Left side: Atomic configuration for the three-level cascade system in
$^{24}$Mg, utilizing the $(3s^2)$ $^{1}S_{0} \rightarrow (3s3p)$ $^{1}P_{1} \rightarrow (3s3d)$
$^{1}D_{2}$ transition, with levels denoted by $\ket{0}$, $\ket{1}$, and $\ket{2}$, respectively.
The quantum interference between various excitation pathways in this system can be used to increase
the effectiveness of laser cooling. The energy of the lower and upper transition is
$\hbar\omega^{(1)}_{0}$ and $\hbar\omega^{(2)}_{0}$, respectively. The frequency and detuning for
each laser is labeled by $\omega_{i}$ and $\delta_{i}$, respectively, for $i=1$, $2$. The
spontaneous-emission linewidths of the states $\ket{1}$ and $\ket{2}$ are $\Gamma_1$ and
$\Gamma_2$, respectively. Right side: Characteristics of the dressed $^{24}$Mg cascade system, with
$s_1(\delta_1) = 0.001$ and $s_2(\delta_2) = 1$. Real (top graph) and imaginary (bottom graph)
parts of the eigenvalues of the Hamiltonian in Eq.~(\ref{eq:ham}), with dressed atomic states
labeled for the manifold. The real parts are the energies and the imaginary parts are the effective
linewidths of the dressed atomic system. Both are plotted as functions of the dressing
laser detuning $\delta_2$, with fixed $\delta_1 = 0$.}
\end{center}
\end{figure}

This Letter presents a general method that uses coherent quantum interference to tailor an
atom's internal energy structure that makes it more amenable to simple Doppler
cooling in free space. To exploit the effects of quantum interference, at least three internal
atomic levels and two lasers with distinct frequencies must be utilized. There are three basic
types of three-level systems
--- the $\Lambda$-, V-, and $\Xi$-systems --- each classified according to the ordering of the bare
quantum states in energy, and the possible decay pathways. The $\Lambda$-configuration is commonly
used for studying of EIT~\cite{imamoglu_05} and related
phenomena. Here we focus on a $\Xi$-type system, also known as a cascade configuration.  We
note, however, that our general conclusions can be applied to any type of three-level system, and
can be extended to systems with more than three levels. In particular, we will focus our discussion
on a cascade system suggested by Magno et al.~\cite{cruz_03} as a simpler second-stage cooling
scheme for alkaline-earth atoms, which has recently been demonstrated for magnesium~\cite{cruz_05}.
The level structure and internal parameters for this system are depicted in Fig.~\ref{fig:cascade}.
The cooling scheme is more effective for $\Lambda$-type three-level systems having metastable lower
states, but also works for ladder systems with an upper state narrower than the intermediate
one~\cite{cruz_03}. This cooling technique seems well suited for alkaline-earths, which are good
candidates for the next-generation optical atomic clocks, studies of ultracold collisions, optical
Feshbach resonances~\cite{ciurylo2005ots}, and achievement of quantum
degeneracy~\cite{takasu2003ssb}. Other elements with similar structure include Zn, Cd, and Hg. For
lighter alkaline-earth atoms, such as Ca and Mg, for which techniques other than Doppler cooling
are impossible or too difficult to be used~\cite{cruz_03}, it will facilitate trapping in optical
potentials and lattices.
Our treatment solves the fully quantum-mechanical equations of motion for
the three-level atom in the field of both lasers. The dynamics
of this system are determined exactly using numerical methods,
giving quantitative predicted
laser-cooling temperatures. The results of these calculations then provide us with a basis for a
more intuitive interpretation of the cooling mechanism. Along these lines, we develop a
dressed-state picture of the system dynamics that makes qualitative predictions that agree well
with the numerical calculations.

Fig.~\ref{fig:cascade} shows the internal atomic states in order of increasing
energy as $\ket{0}$, $\ket{1}$, and
$\ket{2}$. The transition energy of the lower transition,
$\ket{0} \rightarrow \ket{1}$, and of the upper transition, $\ket{1} \rightarrow \ket{2}$, are
$\hbar\omega^{(1)}_{0}$ and $\hbar\omega^{(2)}_{0}$, respectively. We include two lasers, of
frequency $\omega_1$ and $\omega_2$, and define their detunings from the appropriate atomic
transitions as $\delta_i =  \omega_i - \omega^{(i)}_0$, for $i =$ 1,2. The intensities of lasers 1
and 2 are characterized by their Rabi frequencies $\Omega_{1} = -\braket{0 | \mathbf{d} | 1} \cdot
\mathbf{E}_{1}(\mathbf{x})$ and $\Omega_{2} = -\braket{1 | \mathbf{d} | 2} \cdot
\mathbf{E}_{2}(\mathbf{x})$, respectively, where $\mathbf{d}$ is the electric-dipole operator of
the atom and $\mathbf{E}_{i}$ is the electric-field amplitude for the $i$th laser. The
spontaneous-emission linewidths of states $\ket{1}$ and $\ket{2}$ are $\Gamma_1$ and $\Gamma_2$,
respectively (88 MHz and 2.2 MHz, respectively, for $^{24}$Mg). The time evolution for the atom
moving in the laser field, with mass $m$ and center-of-mass momentum operator $\mathbf{p}$, is
described by the master equation,
\begin{equation}
 \label{eq:master}
 \dot{\rho}(t) = \frac{i}{\hbar} \left[\rho,H\right] + \mathcal{L}\left[\rho\right],
\end{equation}
where $\rho$ is the reduced density matrix of the atom system, the vacuum photon field degrees of
freedom having been traced over, $H = \mathbf{p}^2 / 2m + \hbar\omega^{(1)}_{0}\ket{1}\bra{1} +
\hbar\omega^{(2)}_{0}\ket{2}\bra{2} + V^{(1)}_{\text{laser}}(\mathbf{x}) +
V^{(2)}_{\text{laser}}(\mathbf{x})$ is the Hamiltonian of the atom-laser system, and the
superoperator Liouvillian $\mathcal{L}$ describes effects due to coupling of the atom to the vacuum
photon field, resulting in spontaneous emission~\cite{cohen_tannoudji_book_92}.

Eq.~(\ref{eq:master}) treats all the atomic degrees of freedom (internal and external)
quantum mechanically, so its solutions generally provide an accurate description of the atom's
dynamics. We thus avoid much of the ambiguity and difficulty associated with semiclassical
approximations of the system. For reasonable temperatures, however, the number of numerical basis
states required to directly solve the problem, even in one dimension, is impractical for most
computers. Here we resolve this problem without resorting to Monte Carlo methods by noting that the
matrix for the linear system equivalent to Eq.~(\ref{eq:master}) is very sparse --- i.e., only a
small fraction of its elements are nonzero. The particular
structure shared by Liouvillian operators $\mathcal{L}$ describing relaxation processes
simplify the numerics. Starting from
the microscopic properties of atomic operators comprising this so-called Lindblad form of
$\mathcal{L}$, we construct a matrix with the zero elements eliminated. The steady-state
solution of Eq.~(\ref{eq:master}) is then found using standard sparse-matrix inversion or
propagation techniques. The result is an exact direct solution of a fully-quantized master
equation.

The steady-state density matrix has been calculated this way in one dimension. The
average kinetic energy is then $\left<p^2/2m\right> = \text{Tr} (\rho\, p^2/2m)$, which we
equate with $\frac{1}{2}k_B T$, where $T$ is the temperature and $k_B$ is Boltzmann's constant.
The large parameter space of the three-level atom two-laser problem has been explored, e.g.,
varying the detunings, $\delta_1$ and $\delta_2$, as well as the strengths
$\Omega_1$ and $\Omega_2$ of the two lasers. However, we find that the laser
strengths are better characterized in terms of how strongly they dress the atom.
The relevant non-resonant saturation
parameters, $s_1$ and $s_2$, for the respective transitions, are
\begin{equation}
s_i = \frac{1}{2}\frac{\Omega_i^2}{\delta_i^2 + (\Gamma_i/2)^2}.
\end{equation}

\begin{figure}
\begin{center}
\includegraphics[width=2.7in]{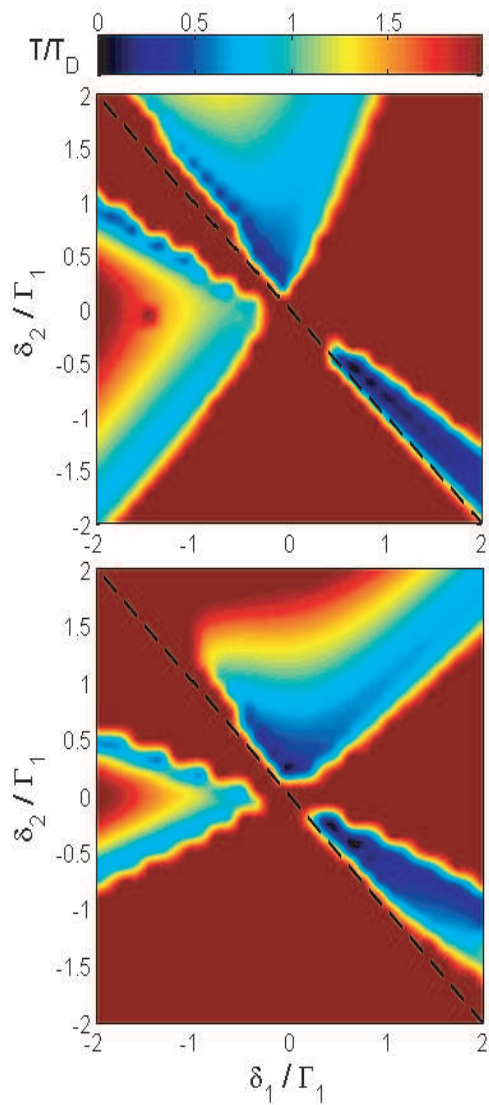}
\caption{\label{fig:temps}(Color online) Steady-state laser-cooling temperatures for a $^{24}$Mg
three-level cascade system, as a function of the two detunings $\delta_1$ and $\delta_2$, obtained
from exact numerical solutions of Eq.~(\ref{eq:master}). The detuning-dependent saturation
parameter for the lower transition (perturbative probe laser) is $s_1(\delta_1) = 0.001$ for both
plots, and for the upper transition (dressing pump laser) is $s_2(\delta_2) = 1$ and 5 in the upper
and lower plot, respectively . The temperature is normalized to the one-dimensional Doppler limit
for the lower transition $T^{(1D)}_{D} = 7 \hbar\Gamma_{1} / 40 k_{B}$~\cite{dalibard_89}, which is
the optimum temperature expected for cooling with just one laser. The dashed line indicates
the location of the bare two-photon resonance. This example illustrates the main parameter regime
where sub-Doppler cooling occurs. Note that, in order to hold the saturation parameters fixed
as the detuning is varied, the Rabi frequencies $\Omega_i$, for $i =$ 1,2, are continuously
adjusted.}
\end{center}
\end{figure}

The parameter space has three distinct regimes. In the first, with $s_1 \gtrsim 1$ and $s_2$
arbitrary, only heating occurs, as expected from Doppler-cooling theory since the
lower transition is driven strongly. In the second, with $s_1 \ll 1$ and $s_2 \ll 1$,
cooling occurs only to
the Doppler limit for the lower transition $T^{(1D)}_{D} = \hbar \Gamma_1 / 2 k_B$, and only in the
range near $\delta_1 = -\Gamma_1/2$. In this case, laser 2 has no effect, as this amounts to simple
two-level Doppler cooling on the lower transition with laser 1. In the final regime, when $s_1 \ll
1$ and $s_2 \gtrsim 1$, cooling occurs down to substantially below $T^{(1D)}_{D}$. Figure~\ref{fig:temps}
illustrates cooling in this regime for the particular case of $^{24}$Mg mentioned above, with the
temperature normalized to $T^{(1D)}_{D}$, and with the bare two-photon resonance ($\delta_1 +
\delta_2 = 0$) denoted by the dashed line. The steady-state temperature is plotted as a function of
$\delta_1$ and $\delta_2$ for $s_1(\delta_1) = 0.001$ and $s_2(\delta_2) = 1$. Note that the
saturation parameters are being held fixed as the detunings are varied, so that the Rabi
frequencies are being continuously adjusted. We see the lowest temperatures, on the order of
$10^{-2}\,T^{(1D)}_{D}$, in the quadrant with $\delta_1 > 0$ and $\delta_2 < 0$, as well as less
extreme cooling in other regions. Observe that the lowest temperatures are obtained for frequencies
detuned to the blue of the two-photon resonance. This seems counterintuitive, since a red detuning
is usually required in order to have a net decrease of atomic kinetic energy in a photon-scattering
event.
Qualitative understanding of the cooling mechanism emerges from analysis of the simpler Hamiltonian,
\begin{equation}
\label{eq:ham}
H = \frac{\hbar}{2}
\begin{pmatrix}
    0      &             \Omega_1       &     0    \\
\Omega_1   &    -2\delta_1 - i\Gamma_1  & \Omega_2 \\
    0      &             \Omega_2       & -2(\delta_1 + \delta_2) - i\Gamma_2
\end{pmatrix}.
\end{equation}
Its complex eigenvalues have real dressed energies, and imaginary parts giving dressed-state
linewidths --- that is, a measure of the coupling of the dressed states to the photon
vacuum. These dressed
energies and widths are plotted on the right side of Fig.~\ref{fig:cascade} as functions of
$\delta_2$, for the same parameters used in Fig.~\ref{fig:temps}.
The cancellation of one of the widths can be viewed as an EIT effect:
since laser 1 is perturbative while laser 2 strongly dresses the upper transition,
the new eigenstates of the system, denoted $\ket{+}$ and $\ket{-}$,
are well approximated as linear combinations of the bare states
$\ket{1}$ and $\ket{2}$. These states have the modified energies and widths shown in
Fig.~\ref{fig:cascade}. The linewidth modifications can be viewed as a
Fano interference~\cite{fano_61}, in which the dressing laser transitions caused by the
probe laser enable multiple coherent pathways among the bare states. Constructive or destructive
interference respectively increases or decreases the atomic linewidth.
\begin{figure}
\begin{center}
\includegraphics[width=2.5in]{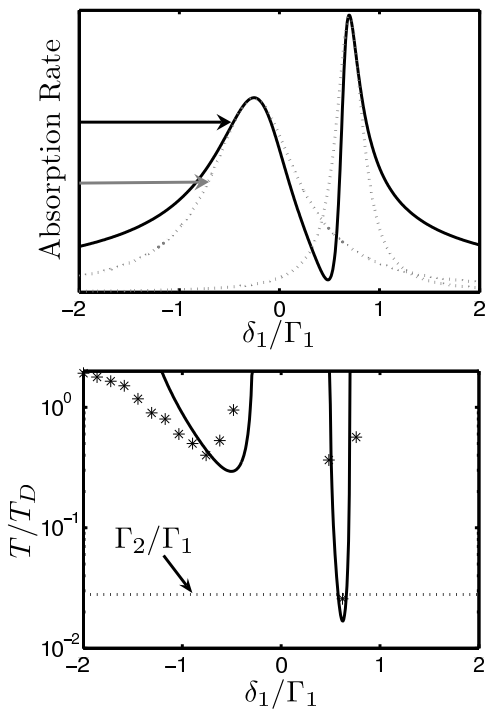}
\caption{\label{fig:lineshape}Upper plot: Absorption spectrum (solid line) as a function of
$\delta_1$, illustrating the asymmetric lineshapes for the dressed system, for a fixed
dressing-laser detuning $\delta_2 = -\Gamma_1$. The peak of each lineshape is located at a dressed
eigen-energy, and lineshapes at the same energies and widths, but with Lorentzian (symmetric)
linshapes are plotted with dotted lines. As an example, optimum-cooling detunings relative to the
leftmost resonance are illustrated with a black arrow for the true asymmetric lineshape and with a
gray arrow for the hypothetical symmetric-lineshape case. The value of the optimum detuning, as
well as the slope of the lineshape is seen to be different for these two cases. Lower plot, solid
line: the ratio of the maximum slope of a Lorentzian lineshape with width $\Gamma_1$ to the slope
of the asymmetric lineshape, as a function of $\delta_1$ with $\delta_2 = -\Gamma_1/2$. This ratio
provides an indication of the expected cooling for the dressed system relative to the Doppler limit
for the lower transition. For comparison, fully quantum numerical results are indicated by data
points.}
\end{center}
\end{figure}

The cooling mechanism is thus qualitatively explained as ordinary two-level Doppler cooling.
But instead of using a transition between two bare states of an atom, the
transition occurs between a (mostly) unmodified ground state, and a dressed excited state, with a shifted
energy and a new linewidth that can be narrower than the bare linewidth of the lower transition. As
the probe laser is scanned, the detuning relative to the dressed energy levels is varied, but since
these levels are shifted from their bare energies, resonance occurs for different detunings
than are encountered in the bare system. In fact, the shifts of the eigen-energies in
Fig.~\ref{fig:cascade} from the bare energies explain the apparent observation of blue two-photon
cooling in Fig.~\ref{fig:temps}. In the dressed system, the bare two-photon resonance is no longer
meaningful, and the cooling region is in fact {\it to the red} of a dressed resonance.

An additional caveat applies when mapping this system onto Doppler cooling theory: the
lineshapes are not Lorentzian, but are in fact asymmetric Fano lineshapes for the dressed system,
as shown in the upper part of Fig.~\ref{fig:lineshape} as a function of $\delta_1$ for a fixed
value of $\delta_2 = -\Gamma_1/2$. This fact changes the optimum-detuning condition into a new
one: maximum cooling for a given transition occurs when the probe laser is detuned from the
dressed excited state {\it precisely to the inflection point of the absorption spectrum}.
This can be understood by
noting that the force $f$ applied to the atom due to the laser beam is proportional to the
absorption rate, for a given $\delta_1$. As is often utilized in semiclassical cooling theories,
the friction coefficient $\alpha$ for the atom moving in the laser field is given by
\begin{equation}
\alpha = -\derfrac{}{v} f(v),
\end{equation}
where $v$ is the atomic velocity. Since the detuning of the laser and the resonant atomic velocity
are linearly related, the derivative of the absorption spectrum with respect to $\delta_1$
also yields a maximum in the cooling force. This is evident in normal Doppler cooling
because the optimum detuning occurs when $\delta = -\Gamma/2$, which is the inflection point of the
Lorentzian. In general then, for asymmetric lineshapes, the optimum detuning does not obey such a
simple relation, but depends on the degree of asymmetry.

From this complete picture of the cooling mechanism as a weak probe applied to a dressed
three-level system, the minimum expected temperatures can now be determined, allowing for
the detuning modification due to asymmetric lineshapes. The lower part of Fig.~\ref{fig:lineshape},
the ratio of the maximum slope of a Lorentzian lineshape with width $\Gamma_1$ to the slope of the
asymmetric lineshape, as a function of $\delta_1$ with $\delta_2 = -\Gamma_1/2$. For comparison,
fully quantum numerical results are indicated by data points. This ratio provides an indication of
the expected cooling for the dressed system relative to the Doppler limit for the lower transition.
Note that the expected temperature, due to the asymmetric lineshape, is predicted to be lower than
the upper-transition Doppler limit, indicated by the horizontal dotted line in the lower plot of
Fig.~\ref{fig:lineshape}. This prediction is supported by the numerical data.

In conclusion, coherent engineering of an atomic three-level system can
optimize the effectiveness of two-level Doppler cooling. By strongly driving a particular
transition between two excited internal states of the atom, dressed energy eigenstates are created
with modified linewidths which, due to mixing, can vary anywhere in the range between the smallest
and the largest of the two bare linewidths. Smaller linewidths lead to lower temperatures, but the
additional effect of asymmetries in the linewidths of the dressed states can lead to even lower
temperatures, below the Doppler limit of the upper transition. The ability to tailor the
degree of cooling lends this technique additional utility. A dressing scheme can be suited to the
characteristics of a particular atom, and real-time adjustment of the cooling properties can allow
narrowing of the velocity-capture range as an atomic gas is cooled. Utilizing such coherent effects
should lead to relatively simple schemes for cooling far below the typical Doppler limit.

We thank N. Andersen  and C. Oates for helpful discussions. J.W.D. and C.H.G. acknowledge support
from the National Science Foundation; J.W.T. acknowledges support from the Calsberg and Lundbeck
Foundation; F.C.C. acknowledges support from FAPESP, CNPq and AFOSR.

\end{document}